\documentclass[amsfonts,nofootinbib,superscriptaddress,aps,twocolumn]{revtex4-1}

\usepackage{amsmath}
\usepackage{amssymb}
\usepackage{appendix}
\usepackage{amsbsy}
\usepackage{bm}
\usepackage{epsfig} 
\usepackage{color}
\usepackage{slashed}
\usepackage{graphics}
\usepackage{subfigure}
\usepackage{graphicx}
\usepackage{lipsum}

\begin{document}
\title{Dark Matter in  a bi-metric universe}
 \author{Carlos Maldonado}
 \email{carlos.maldonados@usach.cl}
\affiliation{Universidad de Santiago de Chile (USACH),
Facultad de Ciencia,
Departamento de  F\'{\i}sica, Chile.}
\author{Fernando M\'endez}
 \email{fernando.mendez@usach.cl}
\affiliation{Universidad de Santiago de Chile (USACH), Facultad de Ciencia,
Departamento de F\'{\i}sica, Chile.}

\begin{abstract}
We study the possibility to describe dark matter in a model of the universe 
with two scale factors and a non-standard Poisson bracket structure characterized by the deformation parameter $\kappa$. The dark
matter evolution is analyzed in the early stages of the universe, and its relic 
density is obtained via the Freeze-In and Freeze-Out mechanism.  We show that  
by fixing $\kappa$ and the  initial
ratio of energy densities present in the different sectors of the universe, the space of  thermal average annihilation cross-sections and dark matter masses compatible with the standard cosmology prior to Big Bang Nucleosynthesis (BBN), is enlarged.  This feature of the model is compatible with non-standard cosmology.

\end{abstract}

\maketitle

\section{Introduction}
Dark matter \cite{Zwicky:1933gu,Freeman:1970mx,Bertone:2016nfn,Salucci:2018} and dark energy \cite{Riess_1998,Perlmutter_1999,Peebles:2002gy,Li:2011sd,Bamba:2012cp} represent about 95\% of the total 
matter-energy  content of the universe \cite{ParticleDataGroup:2020ssz}. 
Even if it is  technically possible to include them in the Einstein 
equations as a source, the origin of such sources remains unclear until 
today.
In other words, we know they exist    and how to incorporate them into 
the model, but a description in terms of fields -- as the 
remaining 4\% of standard model particles -- is still unclear.

On the other hand, the standard cosmological model  rests on the hypothesis of 
isotropic  and homogeneous three-dimensional space. If there are any inhomogeneities, they 
\cite{Sikivie,Vilenkin,Lukas,Durrer:1999na,Flanagan:2000,
Moffat:2005ii,Bolejko:2011jc}, should be smoothed out during inflation  
\cite{Peebles:1999,Linde:2005ht}.  The metric of the 
universe, therefore,  is described by the 
Friedmann-Lemaitre-Robertson-Walker metric, 
namely
\begin{equation}
    ds^2=\,dt^2+a^2(t)\left( \frac{dr^2}{1-k\,r^2}+r^2\,d\Omega^2\right),
\end{equation}
where $a(t)$ is the scale factor, and
$r$ is the radial (dimensionless) coordinate. The constant $k$ is the 
curvature of  spatial  sections, which will be taken $k=0$, according
to present measurements \cite{ParticleDataGroup:2020ssz}. Finally,
the present time corresponds to $t=0$.

A different scenario has been proposed in a series of papers where 
a universe  with  two scale factors was considered 
\cite{Falomir:2017hwr,Falomir:2018ayx,Falomir_2020, Maldonado:2021aze}. 
These scales might represent two causally disconnected patched 
\cite{RASOULI2019100269} or
two universes in  a multiverse scenario \cite{Linde:2015edk}. The main 
idea of the model is to introduce a sort of interaction through 
the  deformation of the Poisson bracket structure. 

The possibility to have a non-canonical Poisson bracket structure
and the non-commutative algebras associated
\cite{Bayen:1977ha,Fedosov:1994zz,deWilde1983, 
Kontsevich:1997vb,Cattaneo:2004yt} are the source of 
a kind of interaction \cite{Nair:2000ii, 
Gamboa:2001fg,Acatrinei:2001wa,Karabali:2001te,Acatrinei:2001wa}  whose
implications for the dark matter production are investigated in the 
present paper.  Recently,  Poisson bracket deformations
have also been studied  in the context of closed strings 
\cite{Freidel:2017nhg,Freidel:2017wst} and metaparticles 
\cite{Freidel:2018apz}, while the implications for cosmology were discussed in 
\cite{Berglund:2019ctg,Berglund:2019yjq,Berglund:2020qcu,Freidel:2021wpl}.

In the present approach, we  describe the universe by 
two scale factors $a,b$, and the Hamiltonian 
\begin{eqnarray}
\label{ham1}
H&=&\frac{NG}{2} \left[\frac{\pi_a^2}{a}+\frac{1}{G^2}\left(a\,k_a-\frac{\Lambda_a}{3}a^3\right)\right]+
\nonumber
\\
&&
\frac{NG}{2} \left[\frac{\pi_b^2}{b}+\frac{1}{G^2}\left(b\,k_b -\frac{\Lambda_b}{3}b^3\right)\right],
\\
&\equiv&H_a+H_b,
\end{eqnarray}
where $\pi_a,\pi_b$ are the conjugate momenta of $a$ and $b$, respectively
\footnote{The canonical dimension of scale factors is +1.}. $N$ is an auxiliary field 
(chosen $N=1$ at the end of calculations)
 related to the time invariance reparametrization. The  spatial curvatures  
$k_a$, $k_b$ will be set to zero. Finally, the $\Lambda_a, \Lambda_b$, are the  cosmological constants of each patch. 

The Poison bracket structure that introduces the interaction 
between the patches is 
\begin{equation}
\label{pb}
\{a_\alpha,a_\beta\} =0, \quad \{a_\alpha,\pi_\beta\}=\delta_{\alpha\beta},\quad 
\{\pi_\alpha,\pi_\beta\}=\theta\,\epsilon_{\alpha\beta}
\end{equation}
with $\theta$ a constant parameter and  $\{\alpha,\beta\}\in\ \{a,b\}$. In what follows we will use the dimensionless parameter $\kappa$ as the deformation parameter, defined as
$\theta = \kappa\,G^{-1}$.

The field equations derived from this Hamiltonian system are first 
order. They can be recast as the following set of second order
equations
\begin{eqnarray}
\label{eom2a}
2a\ddot{a}+\dot{a}^2 &=&\Lambda_a\,a^2-k_a+2\kappa\dot{b},
\\
\label{eom2b}
2b\ddot{b}+\dot{b}^2 &=&\Lambda_b\,b^2-k_b-2\kappa\dot{a},
\end{eqnarray}
along with the first order constraint
\begin{equation}
\label{const2}
a\dot{a}^2+b\dot{b}^2=\frac{\Lambda_a}{3}\,a^3-k_a\,a+
\frac{\Lambda_b}{3}\,b^3-k_b\,b.
\end{equation}

The physical implications of this model for inflation have been discussed
in \cite{Falomir:2017hwr}. Different Poisson bracket deformations, 
including a non-trivial bracket between scale factors, have been analyzed 
in \cite{Falomir:2018ayx,Falomir_2020}.

In a recent study \cite{Maldonado:2021aze}, the effects of including
matter  in the model have been considered. To do that, we assumed
 no interaction between matter evolving on  each patch and
the matter-energy content described through a barotropic fluid.

\begin{figure}[h!]
\centering
\includegraphics[width=0.4\textwidth]{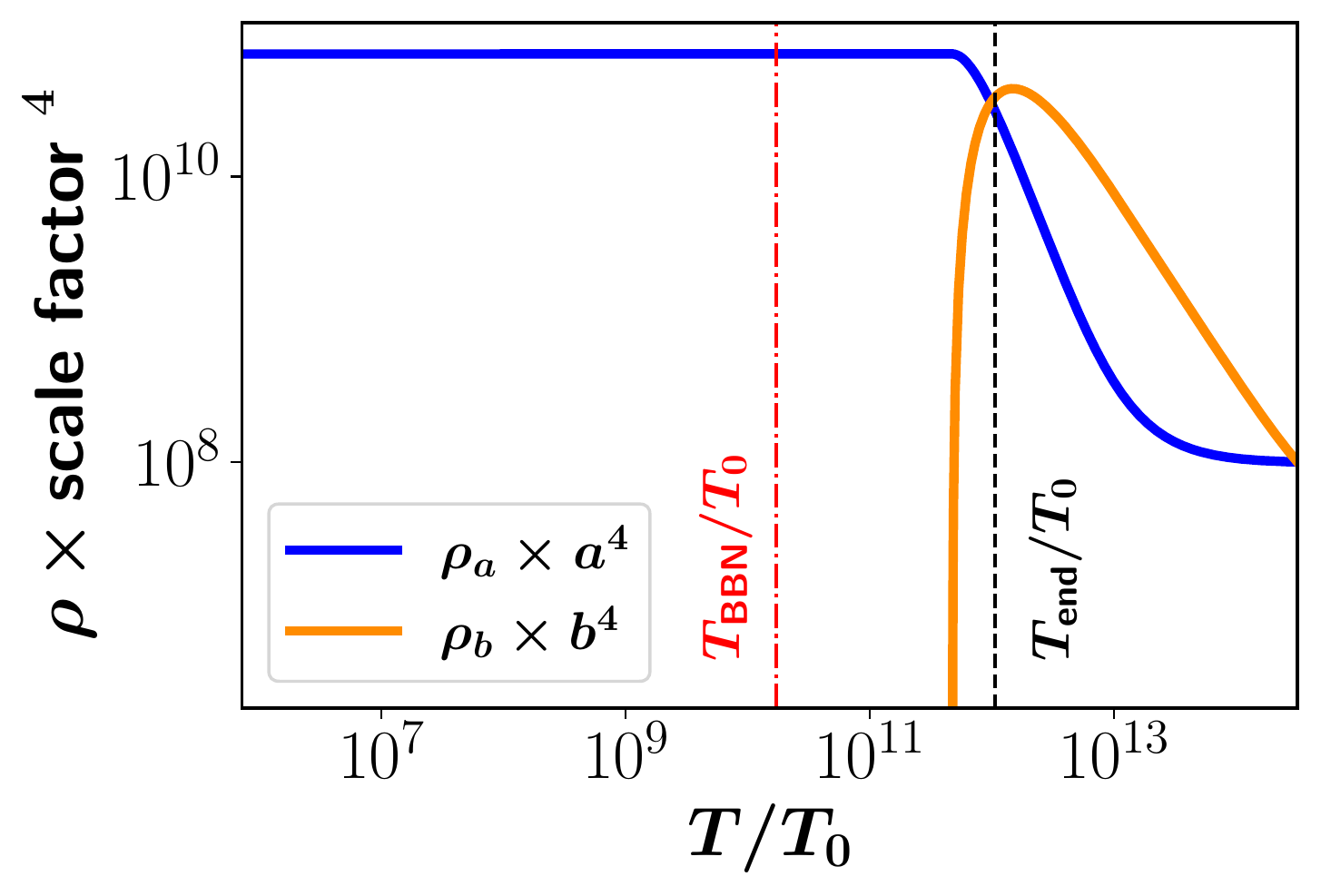}
\caption{Energy densities for $\omega=0$, $\kappa=10^{-35}$ and $\delta=1$. The dashed line represents the temperature at which the field on patch b is no longer effective ($T_\textbf{end}$).}
\label{fig1}
\end{figure}

The numerical analysis showed  a {\it source-sink} effect, that is,  the energy content of one patch {\it drains} to
the other, increasing the energy there. If the process ends before the 
Big Bang Nucleosynthesis (BBN) epoch ($T_\text{BBN}\simeq 4$ GeV), in order not to  conflict with 
astrophysical data \cite{Chung:1999, Kolb:2003}, the standard
cosmology is recovered. This dynamic is described in Figure \ref{fig1}.

The evolution of matter (relativistic
in $a$ and non-relativistic in $b$) is shown as a  function of the ratio $T/T_0$, 
with $T_0$ the present temperature of the universe. The two vertical lines, one at 
$T_{\mbox{\tiny{end}}}/T_0$, and the other at
$T_{\mbox{\tiny{BBN}}}/T_0$, denote temperatures at which the field in $b$ is no longer 
effective (all the energy was drained to patch $a$), and the temperature of BBN, respectively.
In this particular
case, the total drain happens at  the \lq right\rq\, moment of the evolution of patch $a$.

The dynamics previously described turn out to be consistent with a 
kind of  Non-Standard Cosmology scenario \cite{Giudice:2001,Visinelli:2018, 
Maldonado:2019,Arias:2019,Bernal:2019}, in particular with the model where a scalar field is 
introduced to modify the expansion rate of the universe.

In the present research,  we extend the study of matter evolution  to incorporate dark matter (in 
one of the patches), considering an approach that is independent of the particle physics model. The paper is organized as follows. In the next
section, we provide a brief review of two mechanisms of dark matter (DM) production in the 
standard model of cosmology. Section III is devoted to the discussion
of the relevant equations describing DM in the present model. In  section
IV we present the numerical results and the analysis of the 
parameter space consistent with actual observations. The discussion
and conclusions are presented in the last section.

\section{Dark Matter in $\Lambda$CDM}

The Standard Cosmological Model assumes that DM is established in a radiation domination era.  We will focus on two groups of candidates for DM: the Weak Interacting Massive Particles (WIMPs) and Feebly Interacting Massive Particles (FIMPs). The main difference between these two groups is the mechanism of production. WIMPs \cite{Bertone:2004pz, Arcadi:2017kky, Lin:2019uvt, Hooper:2018kfv} are thermally produced via the Freeze-out mechanism \cite{Kolb:1990vq}, and FIMPs \cite{Hall:2009bx, Chu:2011be, Bernal:2017} are generated in a non-thermal mechanism like the Freeze-in \cite{McDonald:2001vt, Choi:2005vq, Kusenko:2006rh}.
The principal characteristics of these mechanisms will be outlined in what follows,
emphasizing the 
aspects  relevant to our proposal.

The  number density of  DM particles, $n_{\text{\tiny{{DM}}}}$, satisfies the Boltzmann equation
\begin{equation}
    \frac{d n_\text{\tiny{DM}}}{dt}+3H\,n_{\text{\tiny{DM}}}=-\langle\sigma v\rangle\left(n_\text{\tiny{DM}}^2-n_{\text{eq}}^2\right),
    \label{DMeq}
\end{equation}
with $H=\dot{a}/a$ the Hubble parameter, $\langle\sigma v\rangle$ the thermal average annihilation cross-section
and $n_\text{eq}$ is the DM  number density of equilibrium.

In the FO mechanism, the DM particles  are  in thermal equilibrium with the bath of particles 
in the early universe,  and as long as the universe expands, their interactions become inefficient 
to maintain  the thermal equilibrium. Therefore, DM particles leave the thermal bath and freeze their number. This  process is referred to as  Freeze-Out. 

The mechanism is described by  eq.~(\ref{DMeq}) and the analytic solution can be obtained in the limit $n_\text{\tiny{DM}}\gg 
n_\text{eq}$.  In this case, the  Yield ($Y$) of DM -- defined as
$Y\equiv n_\text{DM}/s$, with $s$  the entropy density of the universe --  can be estimated as
\begin{equation}
    Y\propto \frac{1}{M_{DM} J\left(x_\text{fo}\right)},
\end{equation}
where $M_{DM}$ is the mass of the DM particle,
$J\left(x_\text{fo}\right)=\int_{x_\text{fo}}^\infty \,x^{-2}\langle\sigma v\rangle (x) dx $, 
with $x$ a dimensionless parameter defined by $x=M_{DM}/T$. The constant  
$x_\text{fo}$ is the  moment at which the DM particle leaves the thermal 
bath. Note that for constant thermal average annihilation cross-section,  the integral turns out to be
$\langle\sigma v\rangle/x_\text{fo}$.

\begin{figure}[h!]
\centering
\includegraphics[width=0.4\textwidth]{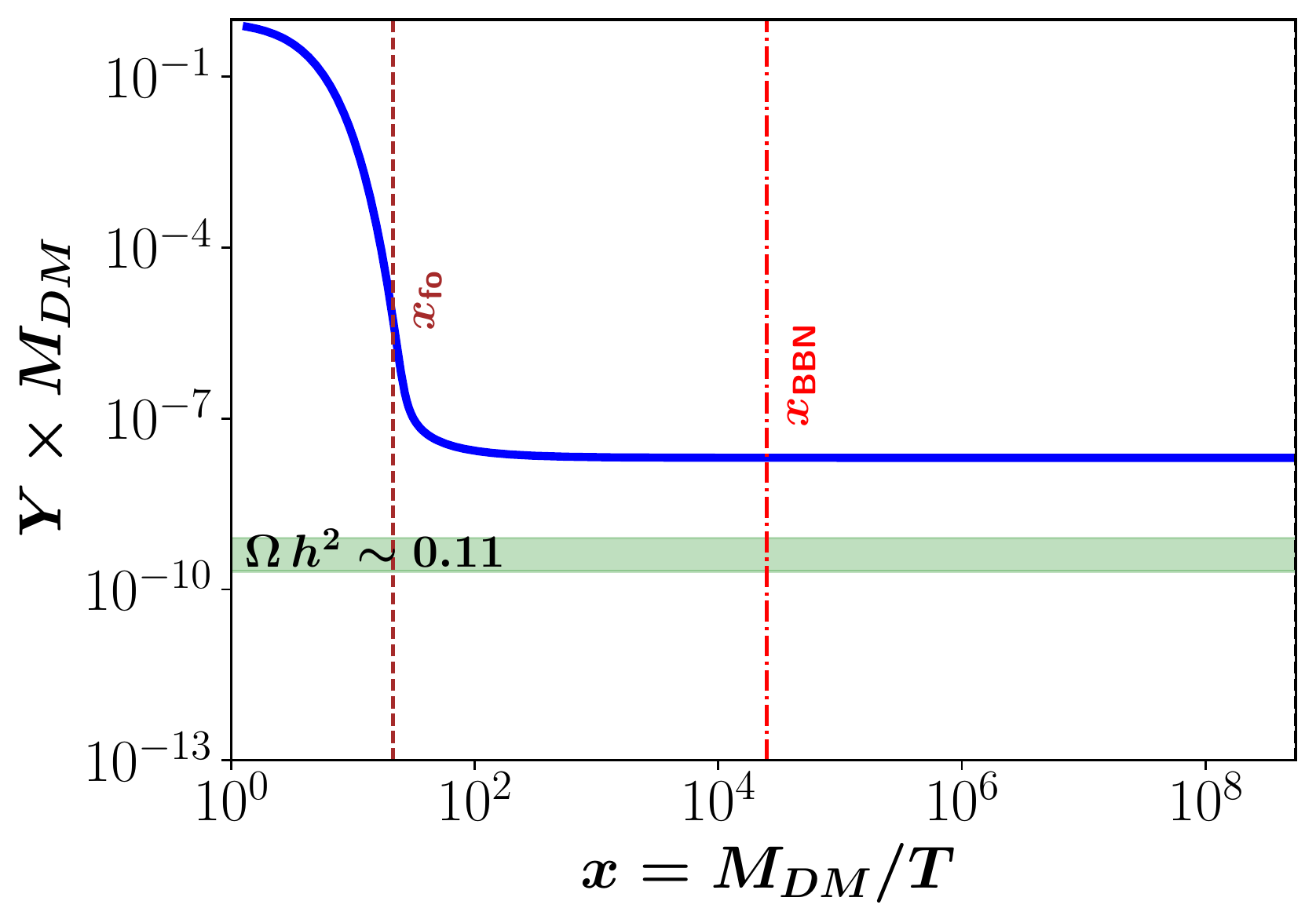}
\caption{Yield of Dark Matter particle in Standard Cosmology with mass $M_{DM}=100$~GeV and thermal average annihilation cross-section $\langle\sigma v\rangle= 10^{-11}$~GeV$^{-2}$ established in the Freeze-Out mechanism.}
\label{FOSM}
\end{figure}

Figure \ref{FOSM} shows the Yield of DM in terms of $x$ with
$M_{DM}=100$~GeV and 
$\langle\sigma v\rangle=10^{-11}$~GeV$^{-2}$. The dashed line 
corresponds to  the  temperature at which the DM particles decouple from the 
thermal bath and freeze their number ($x_\text{fo}$). The temperature at which BBN epoch start is marked with a dot-dashed line. Finally,  the horizontal  strip represents the current relic
density  of DM. The particular  set of parameters ($M_{DM}$ and $\langle\sigma v\rangle$)  in this figure are excluded
in the $\Lambda$CDM model due to the fact that they  overproduce the current DM relic density.

On the other hand, in the FI mechanism the DM particles are produced in a 
non-thermal way. They are not in equilibrium with the thermal bath,  
and therefore,  interactions with  other particles are feeble and 
result in that these particles never thermalize, causing their number to freeze, a
process known as Freeze-In.

This scenario is described  by Eq.(\ref{DMeq}) and the anlytic solution can be obtained in the limit \linebreak
$n_\text{eq}\gg n_\text{\tiny{DM}}$, giving  the
following estimate of $Y$
\begin{equation}
    Y\propto {M_{DM} \langle\sigma v\rangle}.
\end{equation}
That is, the yield is proportional to the thermal average annihilation cross-section, contrasting 
with the previous case.

\begin{figure}[h!]
\centering
\includegraphics[width=0.4\textwidth]{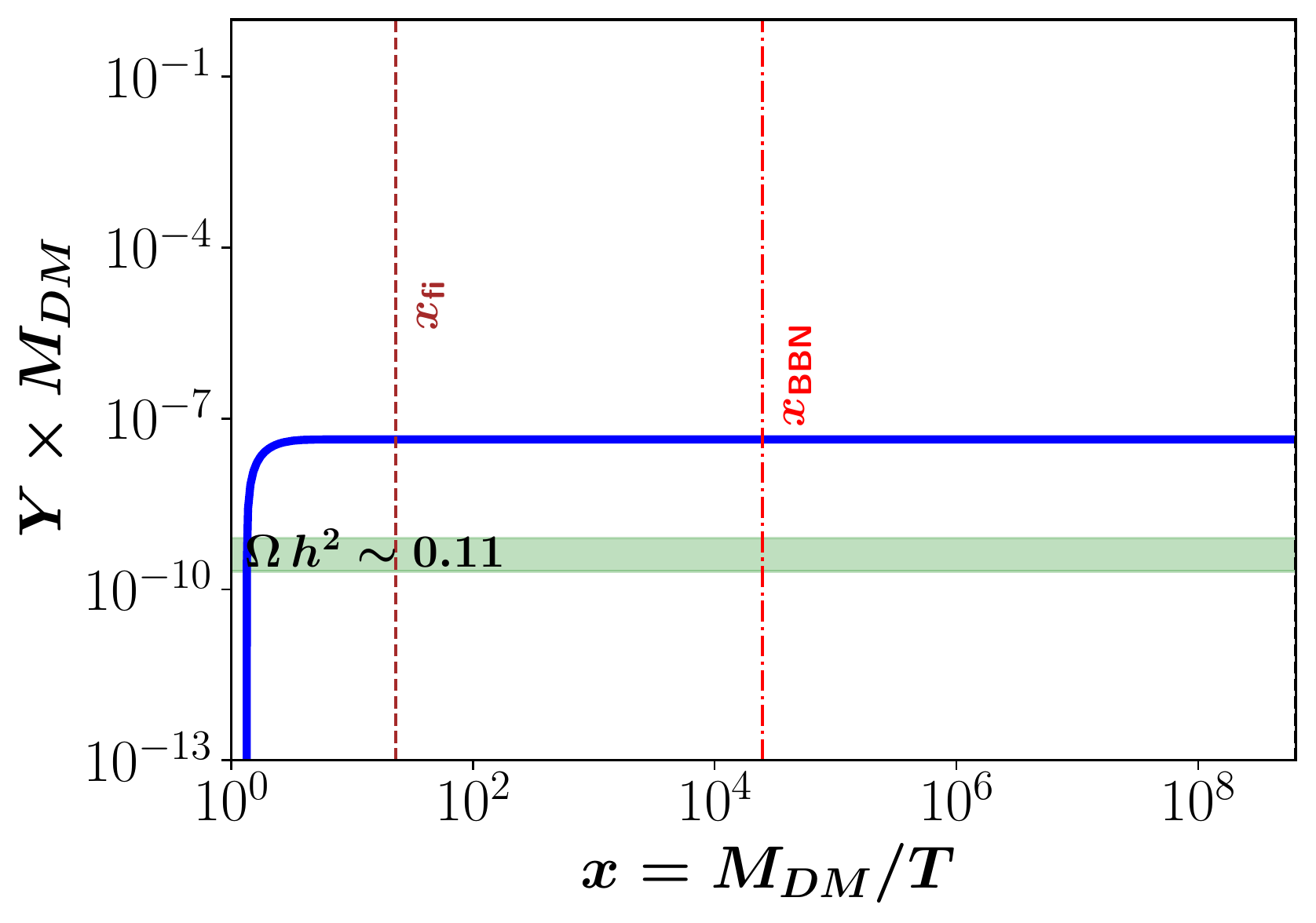}
\caption{Yield of Dark Matter particle in Standard Cosmology with mass $M_{DM}=100$~GeV and thermal average annihilation cross-section $\langle\sigma v\rangle= 2\times10^{-27}$~GeV$^{-2}$ established in the Freeze-In mechanism.}
\label{FISM}
\end{figure}

The main features of this mechanism are depicted in Figure \ref{FISM} with
$M_{DM}=100$~GeV and a thermal average annihilation cross-section 
$\langle\sigma v\rangle=2\times 10^{-21}$~GeV$^{-2}$. The dashed line 
corresponds to the time (temperature)  at which these particles freeze their number ($x_\text{fi}$) while
the horizontal strip is, as before, the current relic density of DM.

{
Like in the FO case, this set of parameters is excluded because the DM relic is overproduce.}

In the following section, we will present the results for both mechanisms in 
the model with two scale factors described in the previous section.

\section{Bimetric universe with matter and Dark Matter}

As anticipated, when the two metrics in presence of  matter are considered, 
the  model behaves as a  of Non-Standard Cosmology, and  the universe 
expands differently  from the standard cosmological scenario. The decaying field in $b$  increases the temperature in the patch  $a$, affecting the production of DM due to 
the entropy density contribution in the Yield.

The field equations for the case when matter is included have been investigated \cite{Maldonado:2021aze}. No interactions between matter content in $a$ with the matter content in $b$ are assumed. Also, the matter is described as a barotropic fluid in both $a$ and $b$.

In the present study, we are interested in the case when  patch $a$ is filled 
with relativistic matter, while the barotropic index $\omega$  characterizes matter in $b$.
 However, for the DM we assume it is present in only one 
patch, which is understood to be the patch  $a$, for definiteness.

Therefore, the set of equations describing the evolution of matter 
density in $a$ and $b$, and DM in $a$ are
\begin{eqnarray}
\dot{\rho_a}+4\frac{\dot{a}}{a} \rho_a &=& 6\kappa M_\text{\tiny{Pl}}^3
\,\frac{\dot{a}\dot{b}}{a^3},
\\
\dot{\rho_b}+3\left(\omega+1\right)\frac{\dot{b}}{b} \rho_b &=& -6\kappa 
M_\text{\tiny{Pl}}^3\, \frac{\dot{a}\dot{b}}{b^3},
\\
\dot{n}_\text{\tiny{DM}}+3\frac{\dot{a}}{a} n_\text{\tiny{DM}}&=&-\langle \sigma v\rangle \left(n_\text{\tiny{DM}}^2-n_\text{eq}^2\right),
\end{eqnarray}
with $\rho$ the energy density (and the subscript denoting the patch),
$M_\text{\tiny{Pl}}$ is the reduced Planck mass, $\kappa$ the deformation parameter and 
$\omega$ the barotropic index of matter in patch $b$ (being $\omega=1/3$ for 
relativistic matter in patch $a$). These equations must consistently solved  together with the 
expressions for the time evolution of the scale factors 
\begin{eqnarray}
\dot{a}&=&a\sqrt{\frac{\rho_a+M_{\tiny{DM}} \,n_\text{\tiny{DM}}}
{3M_\text{\tiny{Pl}}^2}},\\
\dot{b}&=&b\sqrt{\frac{\rho_b}{3M_\text{\tiny{Pl}}^2}}.
\end{eqnarray}
Note that $\rho_{\tiny{DM}}=M_\text{\tiny{DM}}\, n_\text{\tiny{DM}}$
in the evolution equation for $a$. In what follows, the {ratio} between initial
content of matter in $b$ and $a$ will be denoted as $\delta$,
that is $\delta = \frac{\rho_b(M_{DM})}{\rho_a(M_{DM})}$as the most interesting physics occurs at the temperature of DM mass.

In order to investigate this scenario, it is instructive to consider the 
production mechanisms of FO and FI discussed in the previous section. 

For both mechanisms, we  use  same parameters as those in Figure \ref{FOSM} and \ref{FISM} which are
excluded in the $\Lambda$CDM, but  are allowed in the present model. 

\begin{figure}[h!]
\centering
\includegraphics[width=0.4\textwidth]{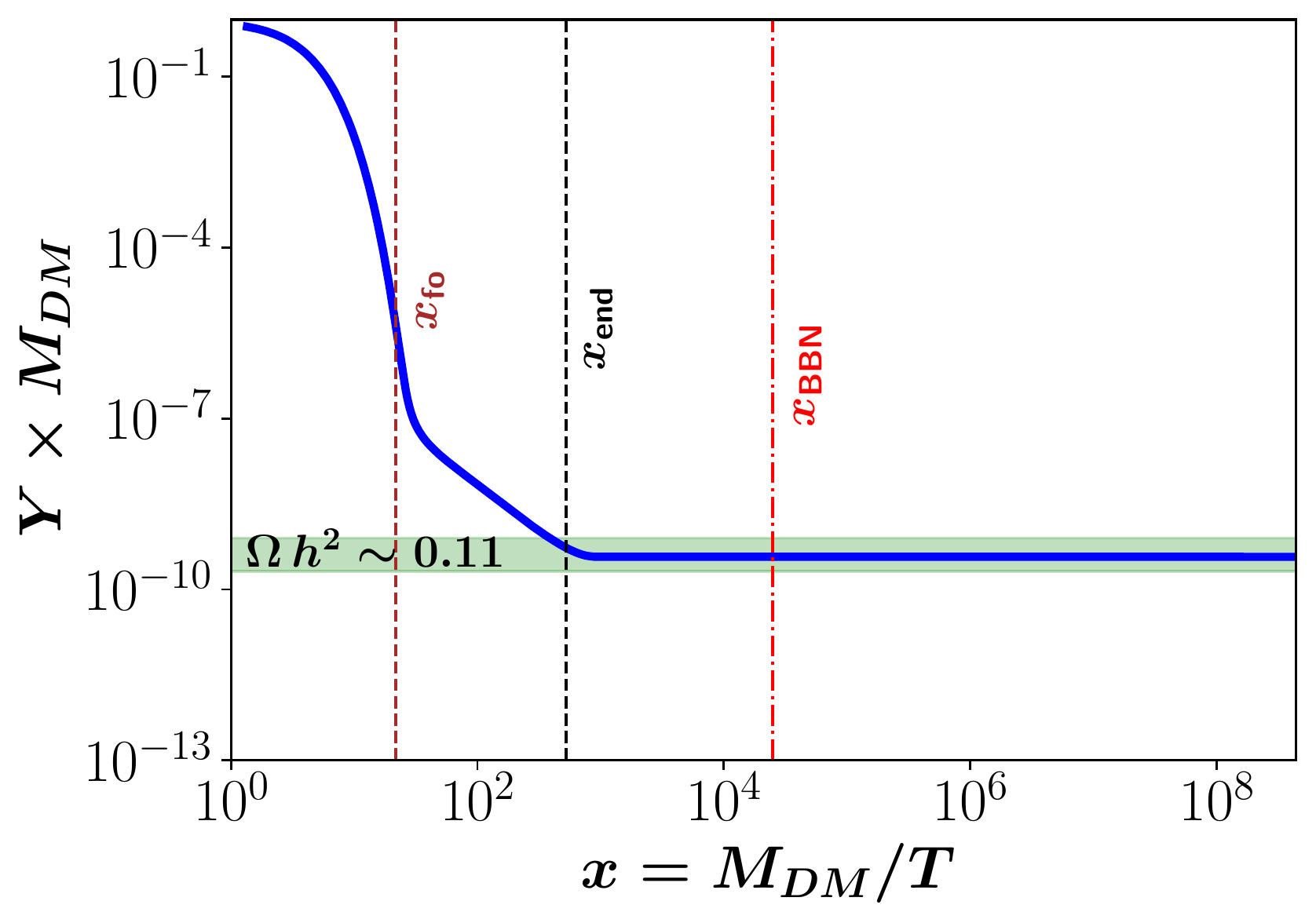}
\caption{Yield of Dark Matter particles with mass $M_{DM}=100$~GeV and thermal average annihilation cross-section $\langle\sigma v\rangle= 10^{-11}$~GeV$^{-2}$ established in the Freeze-Out mechanism in a bimetric universe with $\omega=0$ in $b$ patch, $\kappa=10^{-35}$ and $\delta=1$.}
\label{FONSC}
\end{figure}

We set the deformation
parameter $\kappa=10^{-35}$, with non-relativistic matter in the patch $b$  ($\omega$=0)
and a symmetric initial condition $\delta=1$. The evolution of Yield  can be observed in Figure \ref{FONSC} 
giving us the current relic density with DM parameters that were discarded in the 
$\Lambda$CDM model.

Note also that there are two lines which do not depend
on the model. The first one corresponds to the temperature
at which DM freezes its number (dotted line $x_{fo}$), and
the second is the BBN temperature (dot-dashed  line
$x_{\text{\tiny{BBN}}}$). The line $x_{\text{\tiny{end}}}$, 
corresponds to the moment at which the matter in 
$b$ is no longer effective and, therefore, the Yield 
of DM establishes as a constant number. Finally, 
observe that a condition  $x_{\text{\tiny{end}}}\leq x_{\text{\tiny{BBN}}}$
is required to be consistent with current cosmological data.

\begin{figure}[h!]
\centering
\includegraphics[width=0.4\textwidth]{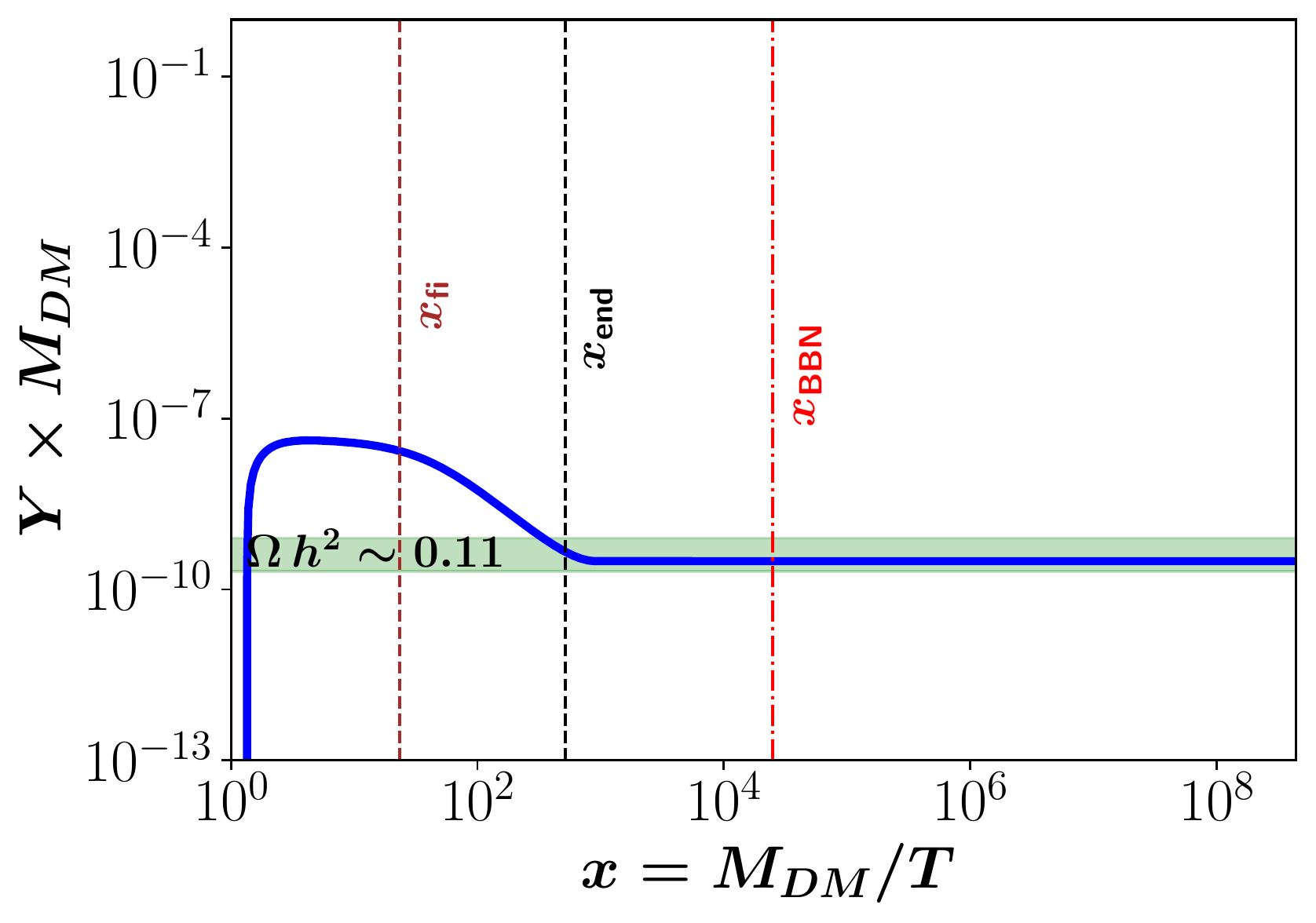}
\caption{Yield of Dark Matter particles with mass $M_{DM}=100$~GeV and thermal average annihilation cross-section $\langle\sigma v\rangle= 2\times10^{-27}$~GeV$^{-2}$ established in the Freeze-In mechanism in a bimetric universe with $\omega=0$ in $b$ patch, $\kappa=10^{-35}$ and $\delta=1$.}
\label{FINSC}
\end{figure}

For the FI mechanism, we set  mass and thermal average annihilation cross-section as done in  Figure \ref{FISM}.  The
deformation parameter is chosen as before,
$\kappa=10^{-35}$, and also the  matter fluid in $b$
is non-relativistic ($\omega=0$) with a symmetric 
initial condition  $\delta=1$. The evolution of the 
Yield is depicted in Figure \ref{FINSC}.

We can observe the same behavior described above. The time at which the DM freezes its number is the same as the Standard Case, but in this model, the Yield of DM decreases due to the decay of the field in $b$, giving us the current relic density of DM.

It is important to mention that the quantity $x_\text{end}$ depends only on the value of $\kappa$ and $\delta$, i.e, for the same value of $\kappa$ and $\delta$, the time at which the effect of field $b$ is no longer effective is the same.

Then, it is  natural to ask if  the values of parameters 
($\langle\sigma v\rangle$, $M_{DM}$, $\kappa$, 
$\delta$, $\omega$) shown before are the only possibility  
consistent with the current DM relic  density. 

\begin{figure}[h!]
\centering
\includegraphics[width=0.4\textwidth]{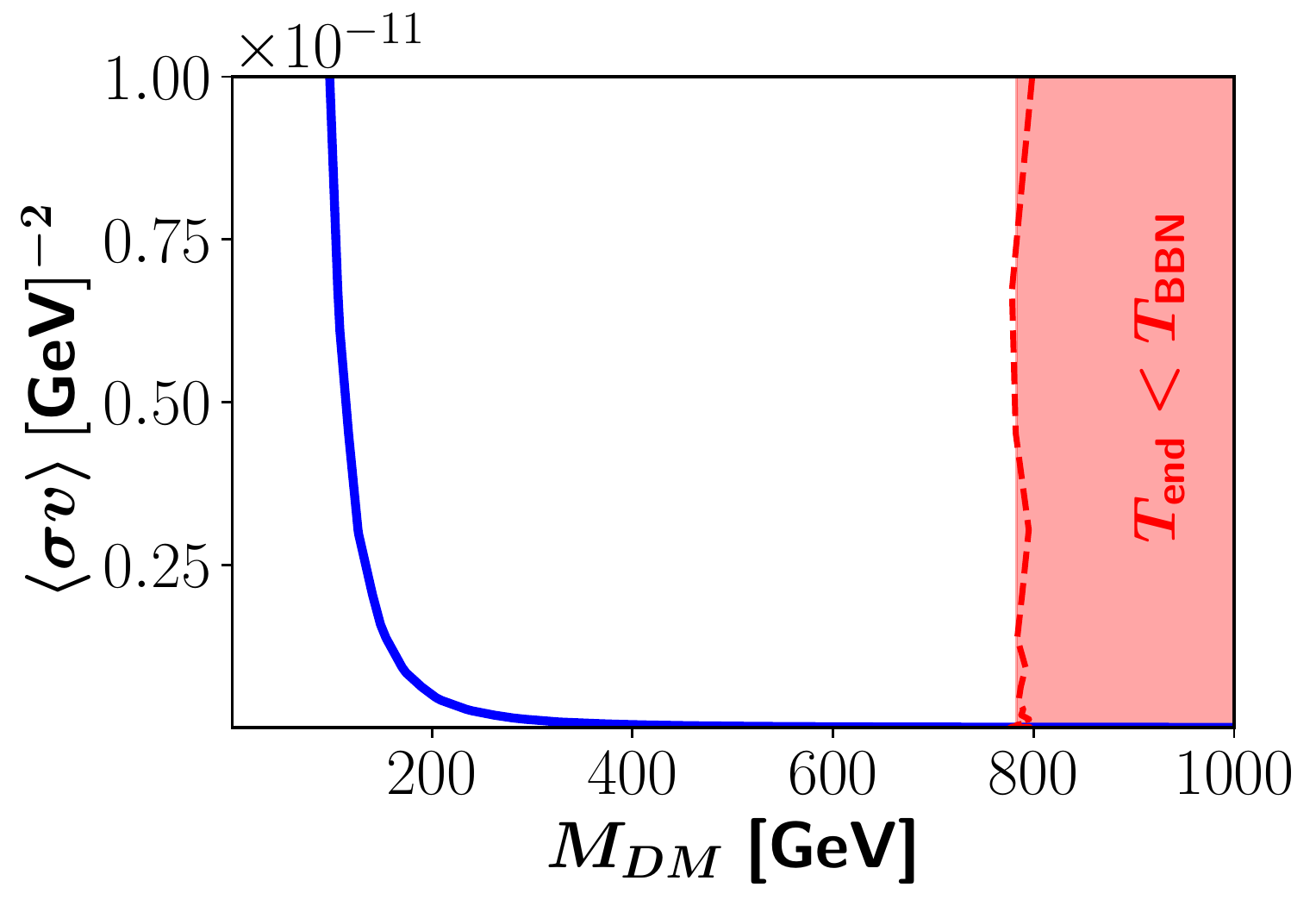}
\caption{Space of parameter in thermal average annihilation cross-section and mass of the Dark Matter particle which gives the current Yield in the Freeze-Out mechanism for $\omega=0$ in $b$ patch, $\kappa=10^{-35}$ and $\delta=1$. The red area corresponds to temperatures below the Big Bang Nucleosynthesis and therefore are forbidden to ensure the correct measurements of the $\Lambda$CDM model.}
\label{SPFO}
\end{figure}

A fast answer is shown in Figure \ref{SPFO}. A 
parameter space  plot for FO mechanism where, 
the horizontal axis 
is the DM mass parameter $M_{DM}$, and
the vertical axis contains the values of 
thermal average annihilation cross section $\langle\sigma v\rangle$. Parameters
$\kappa$ and $\delta$ are fixed $\kappa=10^{-35}$ 
and $\delta=1$. 

The continuous line corresponds to the 
DM mass and thermal average annihilation cross-section, consistent with 
DM relic density  observations.

Finally, the region for which the matter content 
of $b$ decays after BBN time (temperature) is shown 
as the region $T_{\text{\tiny{end}}} < T_{\text{\tiny{BBN}}}$. We call this {\it the forbidden
zone}.

\begin{figure}[h!]
\centering
\includegraphics[width=0.4\textwidth]{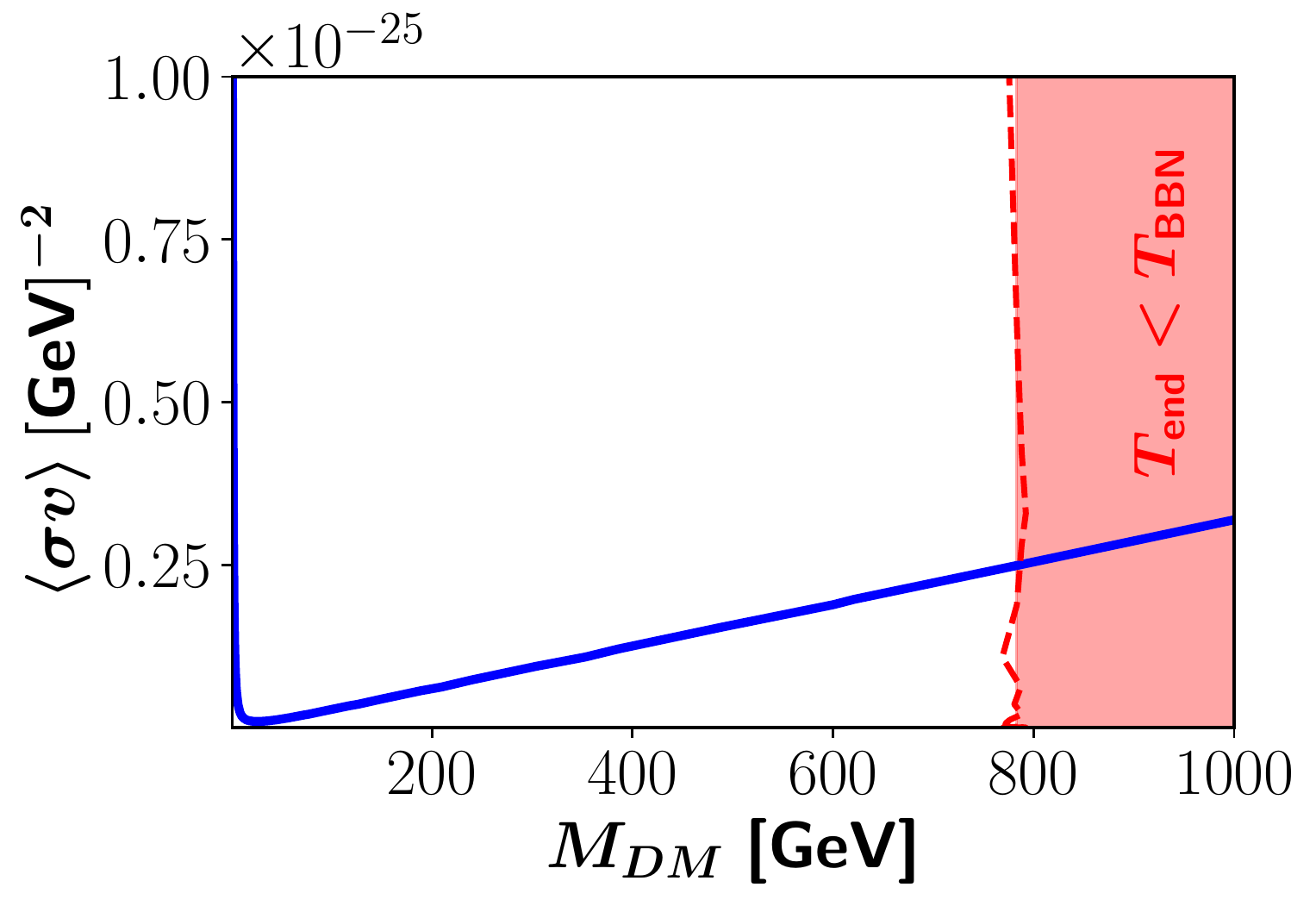}
\caption{Space of parameter in the thermal average annihilation cross-section and mass of the Dark Matter particle which gives the current Yield in the Freeze-In mechanism for $\omega=0$ in $b$ patch, $\kappa=10^{-35}$ and $\delta=1$.The red area corresponds to temperatures below the Big Bang Nucleosynthesis and therefore are forbidden to ensure the correct measurements of the $\Lambda$CDM model.}
\label{SPFI}
\end{figure}

Results for the FI mechanism are shown in Figure  
\ref{SPFI}. The forbidden zone is the same as before since it
depends on $\kappa$ and $\delta$ only. The continuous line
shows the DM mass and thermal average annihilation cross-sections giving the current
values of Yield.

Previous results can be extended for different values
of the deformation parameter $\kappa$, initial condition
$\delta$, and for different content of matter in patch $b$
($\omega\neq 0$).

The numerical results of the study of those scenarios  
are the central topic of the next section.

\section{Parameter Space}

When DM is not present, the values of $\kappa$ compatible with present observations 
are those for which the drain of energy from $b$ to $a$ finish before BBN epoch (see \cite{Maldonado:2019}). 
When DM is considered, non-zero values of  $\kappa$ also affect the production of DM relic density.

In order to investigate these effects, in the following subsections we consider two cases. 
In the first, the matter content of the $b$ sector is  non-relativistic  ($\omega=0$), 
and in the second, we consider $b$ patch filled with relativistic matter ($\omega=1/3$).

For both, we consider the FO and FI mechanisms and also will address the problem of non-symmetric initial configuration.

\subsection{Non-relativistic matter in $b$ patch ($\omega=0$)}

\begin{figure}[h!]
\centering
\includegraphics[width=0.4\textwidth]{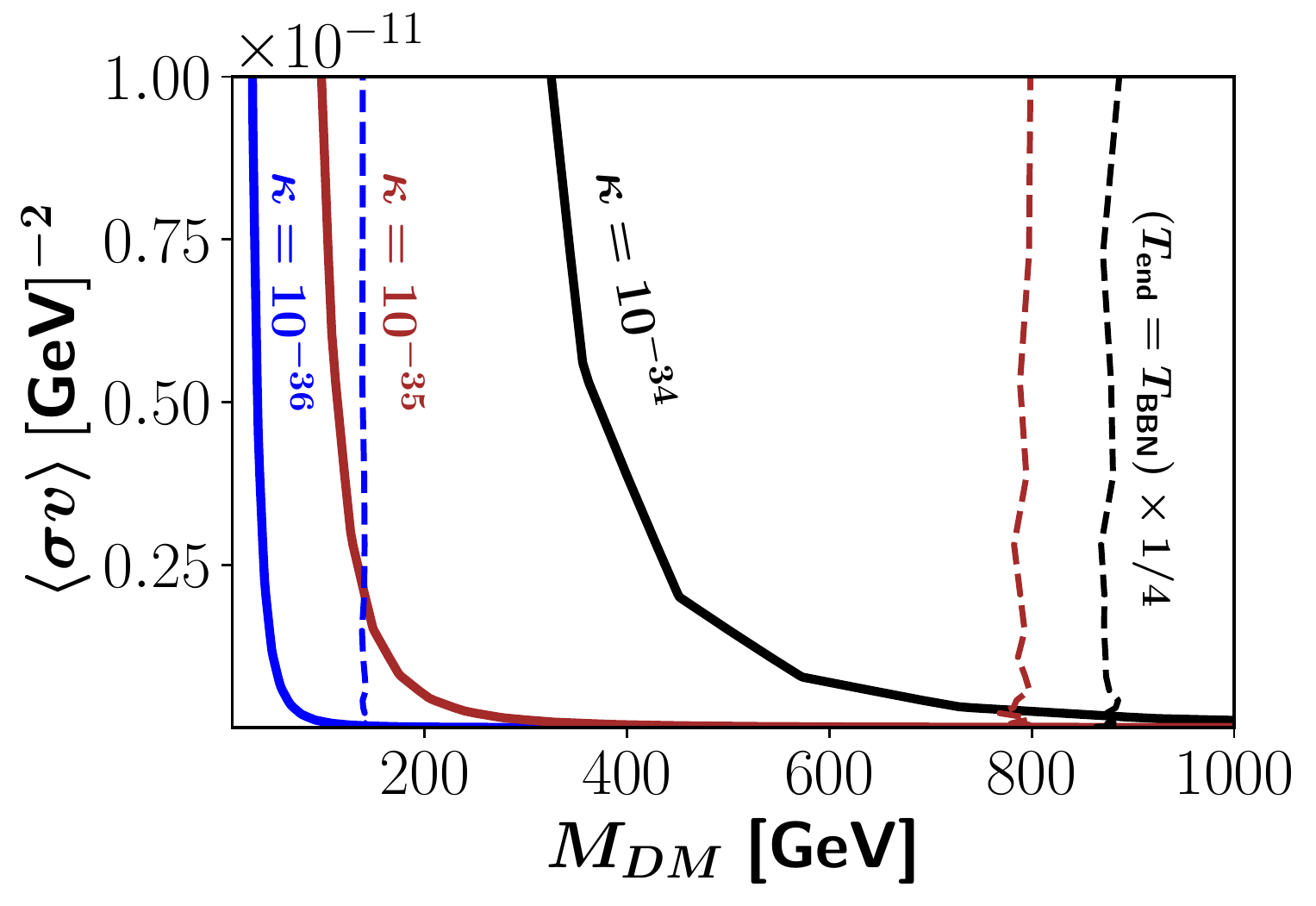}
\caption{Space of parameter in the thermal average annihilation cross-section and mass of the Dark Matter particle which gives the current Yield in the Freeze-Out mechanism for different values of $\kappa$, $\omega=0$ and $\delta=1$. The dashed lines correspond to $T_\text{end}=T_\text{BBN}$. Points  to the right of this line will be in conflict with actual  measurements of the $\Lambda$CDM model. The black dashed line is shifted by a factor of 1/4.}
\label{FOkappa}
\end{figure}

For the FO mechanism and $\delta=1$, the space of allowed DM mass and thermal average annihilation cross-section 
for different values of $\kappa$, 
 are the solid lines shown in  Figure \ref{FOkappa}. The dashed lines (with 
corresponding colors) show the boundary of the forbidden zone so that for all points
to the right of this boundary, $T_\text{end}$ are below of the BBN temperature.

Large values of DM mass with smaller values of thermal average annihilation cross-section are allowed as $\kappa$ 
increases. 

\begin{figure}[h!]
\centering
\includegraphics[width=0.4\textwidth]{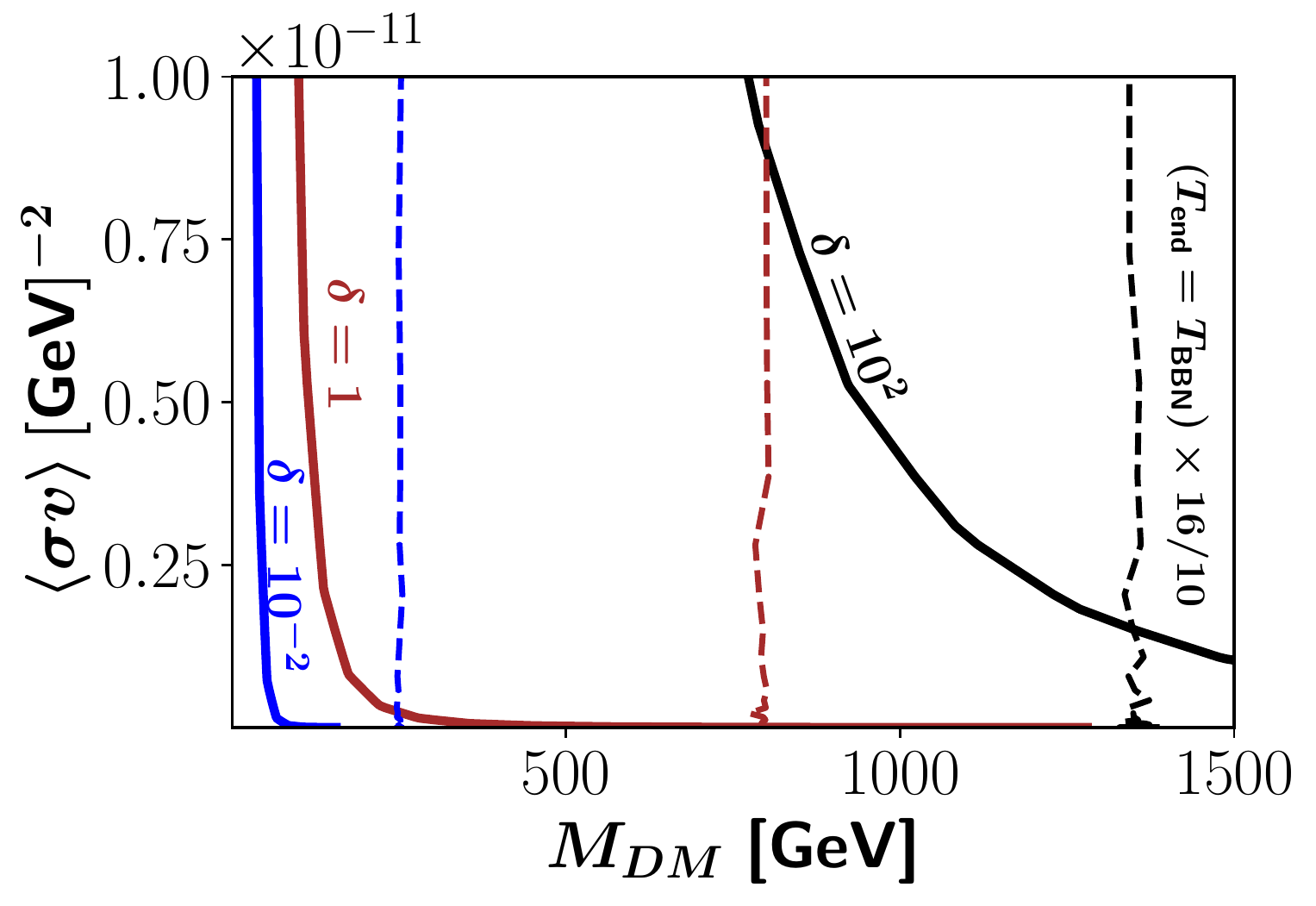}
\caption{Space of parameter in the thermal average annihilation cross-section and mass of the Dark Matter particle which gives the current Yield in the Freeze-Out mechanism for different values of $\delta$, $\omega=0$ and $\kappa=10^{-35}$. The dashed lines correspond to $T_\text{end}=T_\text{BBN}$. Points  to the right of this line will be in conflict with actual  measurements of the $\Lambda$CDM model. The black dashed line is shifted by a factor of 16/10.}
\label{FOdelta}
\end{figure}

Now we consider fixed  $\kappa=10^{-35}$ for variable initial condition $\delta$ in the 
FO mechanism. The Figure 
\ref{FOdelta} shows a similar behavior to the one just described. Dashed lines show the boundary 
of the forbidden regions (one for each value of $\delta$) and we see that for less symmetric
conditions, it is possible to consider higher values of DM mass and smaller values of the thermal average annihilation cross-sections.

\begin{figure}[h!]
\centering
\includegraphics[width=0.4\textwidth]{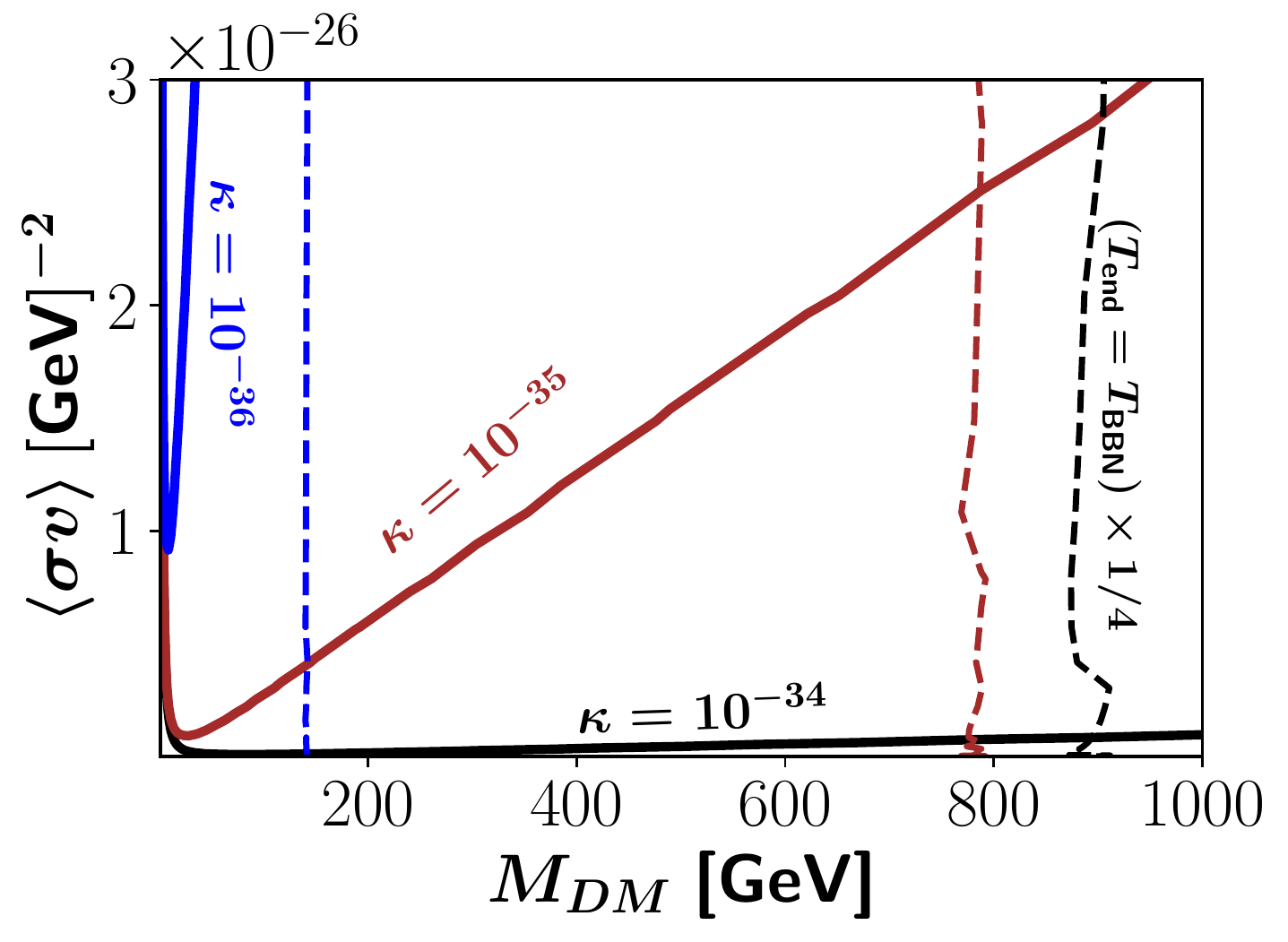}
\caption{Space of parameter in the thermal average annihilation cross-section and mass of the Dark Matter particle which gives the current Yield in the Freeze-In mechanism for different values of $\kappa$, $\omega=0$ and $\delta=1$. The dashed lines correspond to $T_\text{end}=T_\text{BBN}$. 
 Points  to the right of this line will be in conflict with actual  measurements of the $\Lambda$CDM model.  The black dashed line is shifted by a factor of 1/4.}
\label{FIkappa}
\end{figure}

For the FI mechanism,  we have also explored the cases of varying $\kappa$ with symmetric initial
conditions and the opposite one, namely, fixed $\kappa$ and variable initial conditions. The Figure
\ref{FIkappa}, exhibits the allowed parameter space (solid lines) for different values of 
$\kappa$, and $\delta=1$.

An appealing feature of the model is the existence of two values of  DM mass for the same value of the thermal average annihilation cross-section, at least for $\kappa< 10^{-35}$. For  $\kappa\to 0$, the 
graph shows a tendency to collapse the solid line, indicating that the two values of
DM mass should differ by a small amount.  This characteristic deserves a careful analysis exceeding the  scope of the present work, but to be reported in the near future. 

\begin{figure}[h!]
\centering
\includegraphics[width=0.4\textwidth]{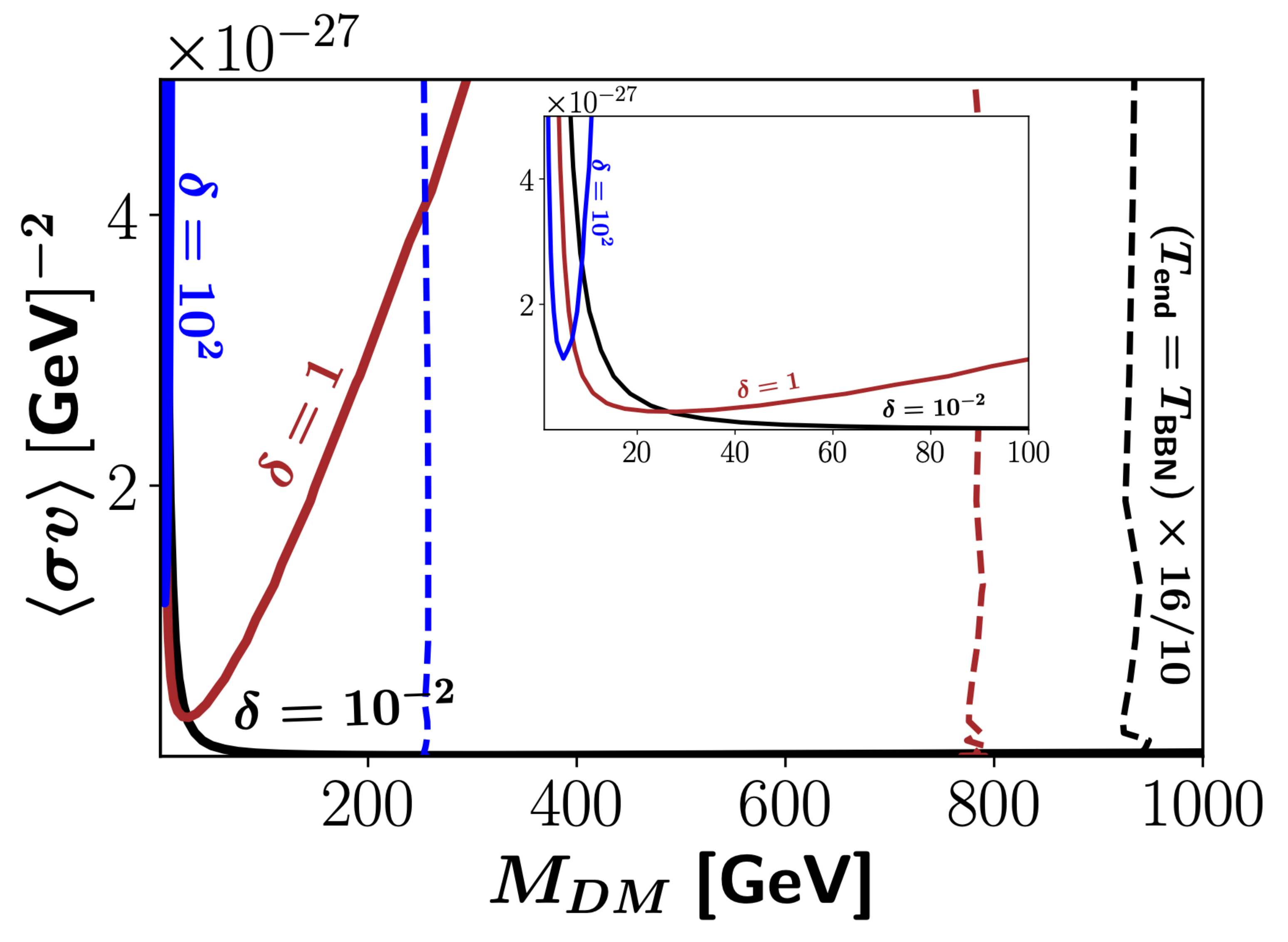}
\caption{Space of parameter in the thermal average annihilation cross-section and mass of the Dark Matter particle which gives the current Yield in the Freeze-In mechanism for different values of $\delta$, $\omega=0$ and $\kappa=10^{-35}$. The dashed lines correspond to $T_\text{end}=T_\text{BBN}$. Points  to the right of this line will be in conflict with actual  measurements of the $\Lambda$CDM model. The black dashed line is shifted by a factor of 16/10.}
\label{FIdelta}
\end{figure}

For the case of variable initial conditions and fixed $\kappa$, results are shown in 
Figure \ref{FIdelta} with $\kappa=10^{-35}$. The inside panel shows a zoom for small
values of DM mass ($M_\text{\tiny{DM}} < 100$ GeV). As before,  higher values in $\delta$  give us branches that open in a wide angle.

Finally note that for the FO mechanism, the allowed parameters (solid lines) 
with  higher values of $\delta$ behaves like allowed parameters for higher values
of $\kappa$. Instead, for the FI mechanism, the higher values of $\delta$ mimic
the behavior of smaller values of $\kappa$.

\subsection{Relativistic matter in $b$ patch ($\omega=1/3$)}

When we consider a relativistic fluid in $b$ patch ($\omega=1/3$) in the FO mechanism,
the possible values of $\kappa$ and $\delta$ giving  the current DM relic density, turn out o be
more restricted.

However, no solution was found for $\kappa$  with $\delta<10^2$,
so that the current DM relic density is be reproduced prior to BBN, in 
the Freeze-Out mechanism.

\begin{figure}[h!]
\centering
\includegraphics[width=0.4\textwidth]{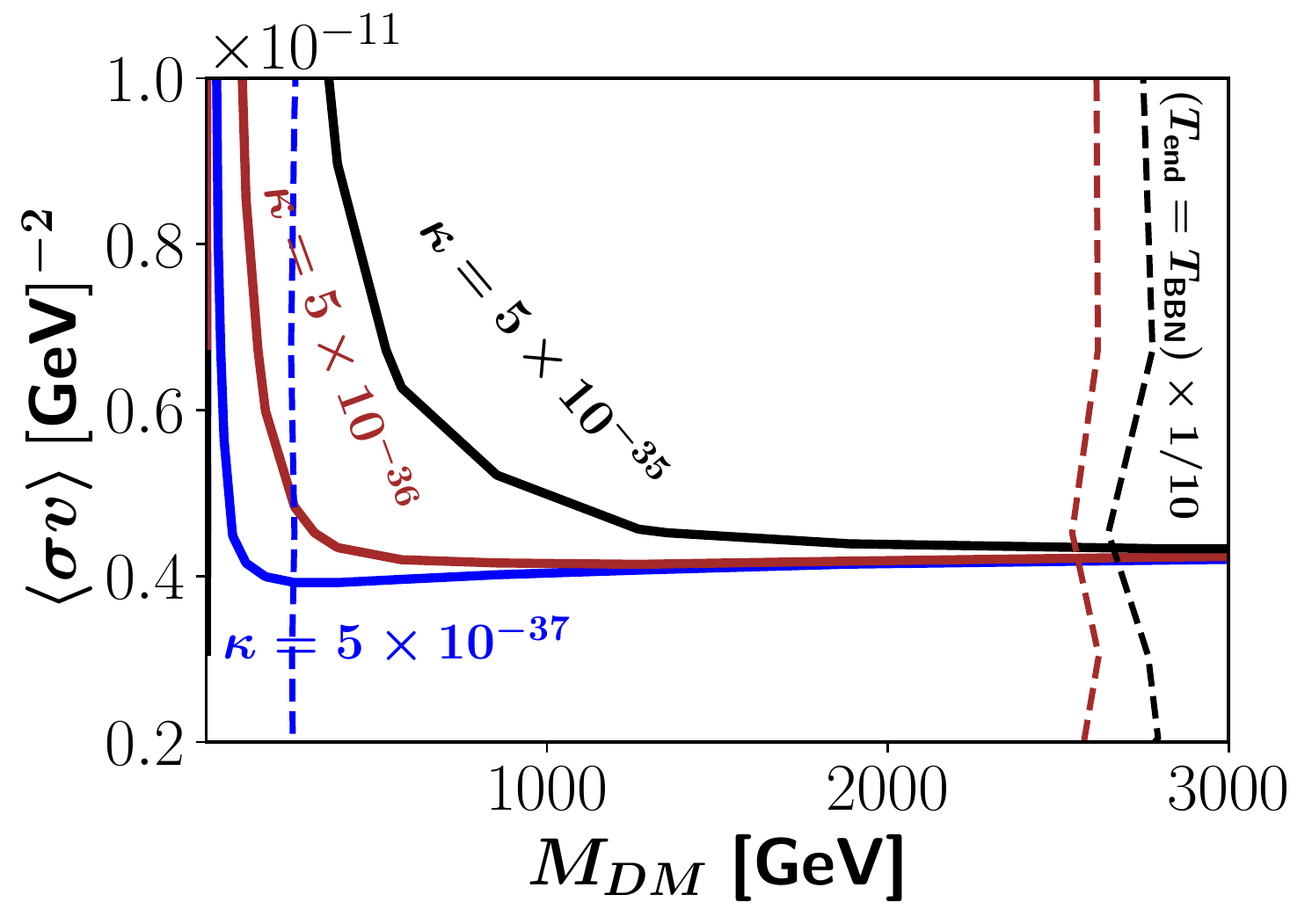}
\caption{Space of parameter in the thermal average annihilation cross-section and mass of the Dark Matter particle which gives the current Yield in the Freeze-Out mechanism for different values of $\kappa$, $\omega=1/3$ and $\delta=10^{3}$. The dashed lines correspond to $T_\text{end}=T_\text{BBN}$. Points  to the right of this line will be in conflict with actual  measurements of the $\Lambda$CDM model. The black dashed line is shifted by a factor of 1/10.}
\label{FOkappaw13}
\end{figure}

The Figure \ref{FOkappaw13} shows different values of $\kappa$ for $\delta=10^3$.  
The boundaries of the forbidden zones are shifted to higher values of DM mass.
Also, we observe that for large values of DM mass, all curves (solid ones)  go to a 
fixed value of the thermal average annihilation cross-section. However,  we emphasize 
that only the  masses to the left of the vertical lines 
(the boundaries of forbidden zones) give  the correct amount of the current  DM relic 
density and, at the same time, $T_\text{end} > T_\text{BBN}$. 

\begin{figure}[h!]
\centering
\includegraphics[width=0.4\textwidth]{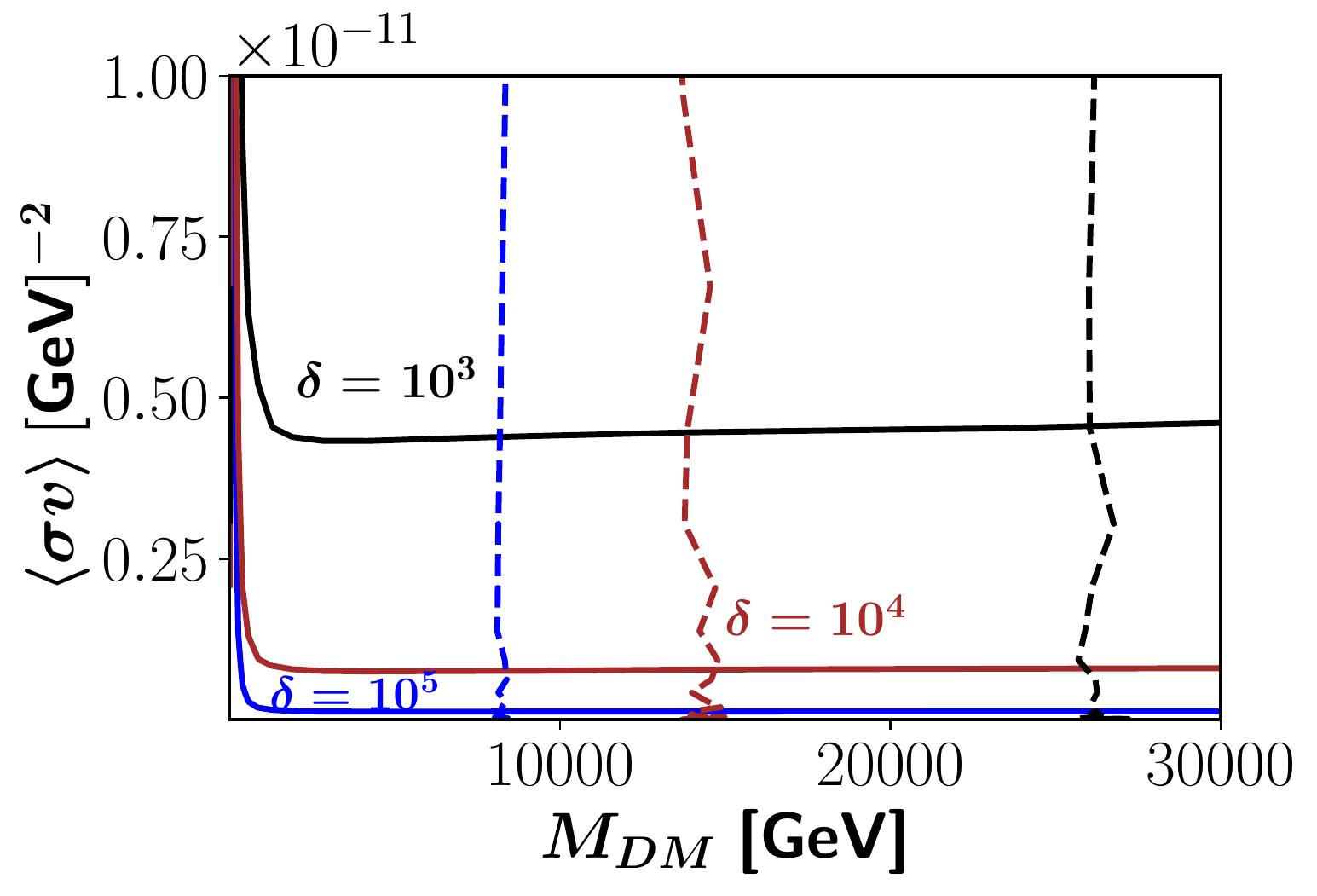}
\caption{Space of parameter in the thermal average annihilation cross-section and mass of the Dark Matter particle which gives the current Yield in the Freeze-Out mechanism for different values of $\delta$, $\omega=1/3$ and $\kappa=5\times 10^{-35}$. The dashed lines correspond to $T_\text{end}=T_\text{BBN}$. Points  to the right of this line will be in conflict with actual  measurements of the $\Lambda$CDM model.}
\label{FOdeltaw13}
\end{figure}

The Figure \ref{FOdeltaw13} shows the parameter space for a constant value of
$\kappa=5\times 10^{-35}$ and different values of $\delta$. In contrast with 
the situation depicted in 
Figure \ref{FOdelta} ($\omega=0$), in this case there is a shift  of the curves  in the y-axis  for higher values of $\delta$. 

\begin{figure}[h!]
\centering
\includegraphics[width=0.4\textwidth]{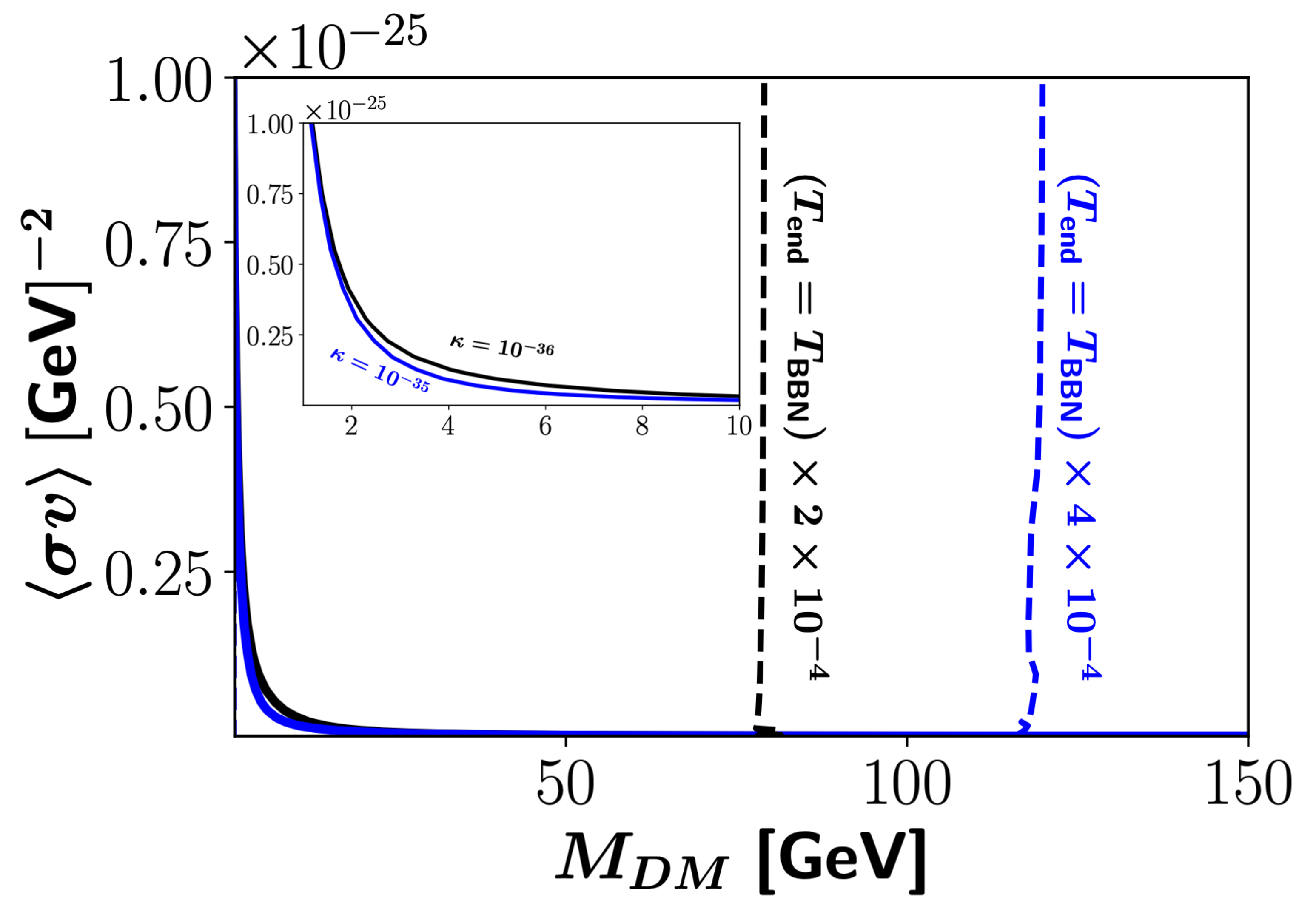}
\caption{Space of parameter in the thermal average annihilation cross-section and mass of the Dark Matter particle which gives the current Yield in the Freeze-In mechanism for different values of $\kappa$, $\omega=1/3$ and $\delta=1$. The dashed lines correspond to $T_\text{end}=T_\text{BBN}$.
 Points  to the right of this line will be in conflict with actual  measurements of the $\Lambda$CDM model.}
\label{FIkappaw13}
\end{figure}

For the FI mechanism the behavior is different from the case of $\omega=0$, but close to
the FO case with $\omega=0$. Therefore, this mechanism allows $\delta <10^2$, in contrast 
with previous one. In Figure \ref{FIkappaw13} we set  $\delta=1$ while $\kappa$ changes,
but only two cases have been considered. High values of $\kappa$ shift the curves 
down in the y-axis, allowing small values of the thermal average annihilation cross-section with high masses of DM. The appearance of 
a minimum is no longer valid.

\begin{figure}[h!]
\centering
\includegraphics[width=0.4\textwidth]{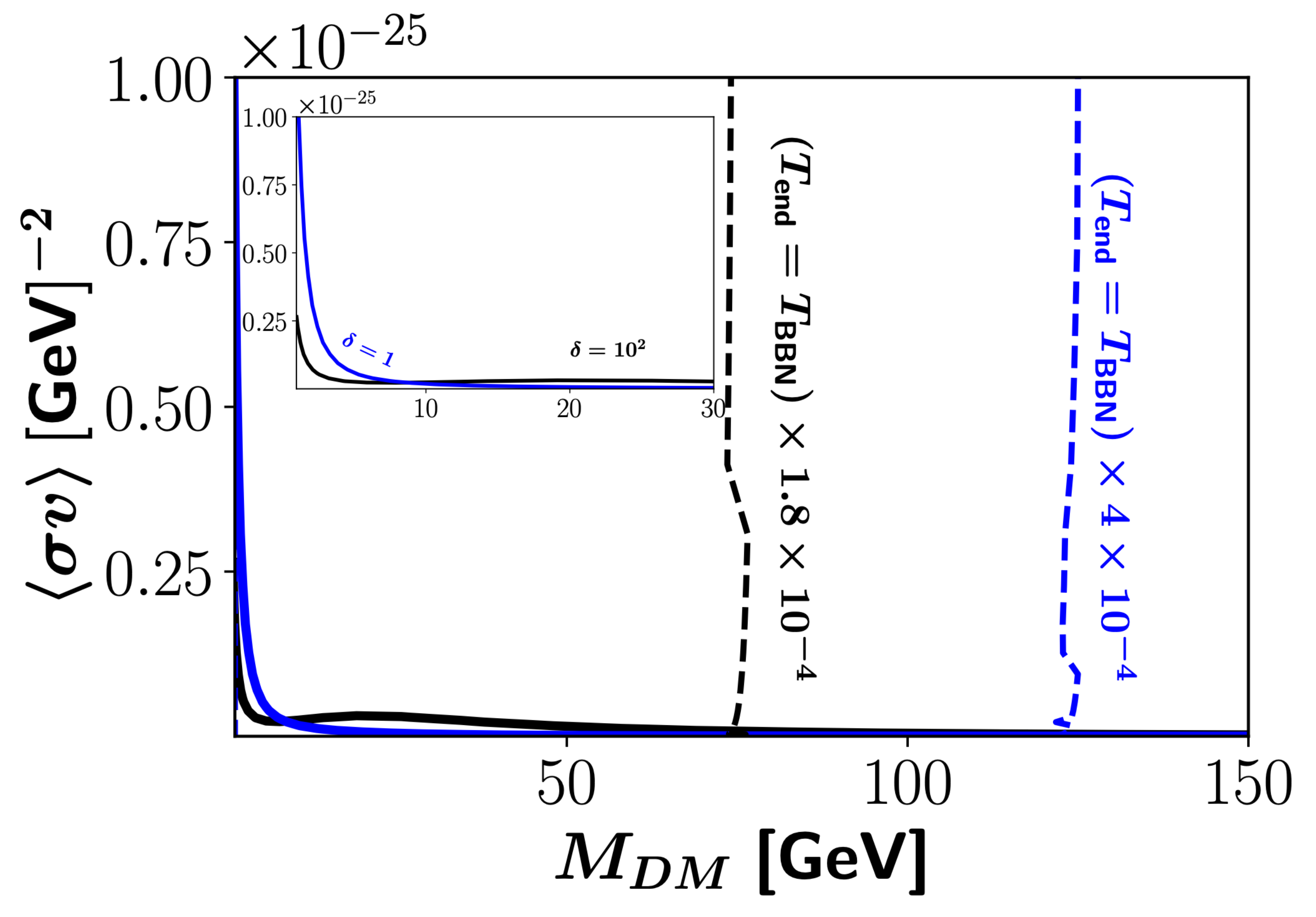}
\caption{Space of parameter in the thermal average annihilation cross-section and mass of the Dark Matter particle which gives the current Yield in the Freeze-In mechanism for different values of $\delta$, $\omega=1/3$ and $\kappa=10^{-35}$. The dashed lines correspond to $T_\text{end}=T_\text{BBN}$. Points  to the right of this line will be in conflict with actual  measurements of the $\Lambda$CDM model.}
\label{FIdeltaw13}
\end{figure}

In the case of a constant value of the deformation parameter ($\kappa=10^{-35}$) and varying $\delta$, 
the allowed parameters are shown in Figure \ref{FIdeltaw13}. As before, no  considerable
differences can be observed for the two values of $\delta$ showed. The forbidden  zone is shifted to
the left for higher  $\delta$, as expected.

\section{Discussion and conclusions}

In this study we have presented the evolution of DM density in a model of the
universe described by two scale factors when there is a non-standard Poisson bracket structure imposed. In a recent study, the evolution of matter in such a model has
been considered and it was shown that the system evolves as a type of non-standard cosmology.

When DM is added to one of the patches  it is natural to ask if  the  
evolution of DM is affected due to the 
evolution of matter in the  other patch present in the model. 

We have shown, by using numerical calculations, that the evolution of DM is indeed affected, but different values of the parameters of the model render such evolution consistent with actual measurement of DM relics. 

In order to do that, two groups of DM candidates were considered: WIMPs and FIMPs, which are related with the mechanism of production Freeze-out (FO) and Freeze-in (FI), respectively. We studied the scenario in which 
DM and relativistic matter coexists in one patch, with the second patch containing 
either relativistic or non-relativistic matter.

For the case when patch $b$ contains non-relativistic matter, we showed that the FO mechanism 
for a symmetric initial condition (the same amount of energy in both patches 
at the beginning of the evolution), higher values of the deformation parameter $\kappa$ allow higher 
DM masses as the thermal avereage annihilation cross-section diminishes. A similar behavior is observed for a constant
$\kappa$, but with initial conditions highly asymmetric ($\delta =10^2$, and higher).

For the FI mechanism, under the same conditions, we observe that there is also a set of DM masses 
and thermal average annihilation cross-sections compatible with the observed actual DM relic. However, in this case, the allowed
values of $\langle\sigma v\rangle$ and $M_{DM}$ have a minimum, either for variable $\kappa$ and symmetric
initial conditions or, for a fixed $\kappa$ and asymmetric initial conditions.

The case for which the $b$-patch is filled with relativistic matter, the possibility of considering
values of thermal average annihilation cross-sections and DM masses different from the usual ones, is also present. Compared to the 
previous case (non-relativistic matter in $b$), here it is possible to reach larger values of  DM masses.

For the FO  mechanism, it was not possible to find a symmetric initial condition compatible with different 
values of $\kappa$ in order to obtain the actual DM relic density prior to BBN. 

The FI mechanism, instead, shows a peculiar behavior. Apart from the fact that the symmetric initial 
condition is admissible, curves with different $\kappa$ (and 
fixed $\delta$), or fixed $\kappa$ (and variable $\delta$), are very close, that is, the range of 
DM masses and thermal average annihilation cross-sections do not  significantly change under variations (independent) of $\kappa$ and $\delta$.

Let us finish our analysis emphasizing the main result: it is possible to consider DM production in this
model in such a way that the DM relic obtained via FI or FO mechanisms is compatible with the actual observations.
The space of parameters of thermal average annihilation cross-sections and DM masses is enlarged, admitting values
which are ruled out in the standard cosmology.

As a final comment, let us point out that in non-standard cosmologies, a similar situation is verified due to the fact that the {\it source-sink} effect is implemented by a $\Gamma\,\rho_\phi$ term,  where $\Gamma$ is the decay rate of a new field ($\phi$) present in the early universe. This $\Gamma$ can be related to the temperature at which the field must decay ($T_\text{end})$. Indeed,
by choosing $M_{DM}=100$~GeV,  $\langle\sigma v\rangle=10^{-11}$~GeV$^{-2}$ and $T_\text{end}=0.1$~GeV, the DM relic density in a FO mechanism is reproduced \cite{Arias:2019} with the same parameter shown in our analysis but from a completely different perspective. In the case of the FI mechanism choosing $M_{DM}=100$~GeV, $\langle\sigma v\rangle=10^{-22}$~GeV$^{-2}$ and $T_\text{end}=500$~GeV  reproduce the DM relic density \cite{Maldonado:2019}, but in our model such a thermal average annihilation cross-section does not reproduce the correct value of DM prior to BBN. This is interesting because for the same values of DM mass it is possible to reach lower values of cross-sections.

\section{Acknowledgements}
This work was supported by Dicyt-USACH grants  USA1956-Dicyt (CM) and Dicyt-041931MF (FM).

\bibliography{biblio}

\end{document}